       \newcommand{\Cc}{ {\mathcal{C}} }

       \newcommand{\Fc}{ {\mathcal{F}} }
       \newcommand{\Gc}{ {\mathcal{G}} }
       \newcommand{\Hc}{ {\mathcal{H}} }

       \newcommand{\Lc}{ {\mathcal{L}} }
       \newcommand{\Mc}{ {\mathcal{M}} }
       \newcommand{\Nc}{ {\mathcal{N}} }

\newcommand{\bB}{\mathbf{B}}
\newcommand{\bv}{\mathbf{v}}

\newcommand{\bJ}{\mathbf{J}}

\newcommand{\nb}{\nabla}

\documentclass[11pt]{epsconf}
\usepackage{graphicx}
\usepackage{wrapfig}
\usepackage{amsmath}
\usepackage{amssymb}
\usepackage{float}
\usepackage{epstopdf}

\title{Certain developments on the equilibrium of magnetized 
plasmas\footnote{Presented at the $45^{th}$ European Physical Society 
Conference on Plasma Physics}}
\author{ A. Evangelias$^1$, D. A. Kaltsas$^1$,  A. Kuiroukidis$^2$,  P. J. Morrison$^3$, G. Poulipoulis$^1$,\\
\underline{G. N. Throumoulopoulos}$^1$}
\institute{$^1$Department of Physics, University of Ioannina, GR 45110 Ioannina, Greece\\ 
$^2$Technological Education Institute of Serres, GR 62124 Serres, Greece\\
$^3$Department of Physics and Institute for Fusion Studies, University of Texas, Austin,\\  Texas 78712, USA}

\begin{document}
\maketitle

%
{\bf Introduction}: 
It has been established in various fusion devices that sheared
flows  play an important role in the transitions to improved
confinement regimes such as the L-H transition and the
formation of internal transport barriers. These flows can
be driven externally with either electromagnetic waves or
neutral beam injection employed for plasma heating and current drive,
or can be created spontaneously (intrinsic flows). Another
important effect of external sources, depending on the
direction of the injected momentum, is pressure anisotropy \cite{Fas, Zwi}, which owing to small collission frequency in  high temperature plasmas is sustained for long time, thus affecting the confinement properties. Also, in large devices as ITER two-fluid effects are expected to become noticeable. 
In the present contribution recent results will be presented on steady states of magnetically confined plasmas obtained by conventional and Hamiltonian methods. The presentation consists   of  three parts. The first one  concerns the derivation of a generalized Grad-Shafranov  (GGS) equation describing helically symmetric equilibria with pressure anisotropy and incompressible  flow of arbitrary direction with application to straight-stellarator  configurations \cite{evku18}. The impact of pressure  anisotropy and flow on the equilibrium characteristics is also examined. In the second part the axisymmetric equilibrium code HELENA  is  extended  for  pressure anisotropy and flow parallel to the magnetic field. 
In the third part the Hamiltonian formulation of helically symmetric plasmas is established within the framework of extended MHD (XMHD), a simplified two-fluid model \cite{kaltsas2017,kaltsas2018}. 

%
{\bf Generalized Grad-Shafranov equation with anisotropic pressure and flow}:
The ideal magnetohydrodynamic equilibrium states with plasma flow and anisotropic pressure are governed by the following system of equations
\vspace{-3mm}
\begin{eqnarray}
\label{momentum}
\varrho(\mathbf{v}\cdot\mathbf{\nabla})\mathbf{v}=\frac{1}{\mu _0}(\mathbf{\nabla}\times \mathbf{B})\times\mathbf{B}-\mathbf{\nabla}\cdot{\rm {\mathbb P}}, \, \, \, \, \mathbf{\nabla}\times (\mathbf{v} \times \mathbf{B})=0, \, \, \, \, \mathbf{\nabla}\cdot (\varrho\mathbf{v})=0, \, \, \, \, \mathbf{\nabla}\cdot\mathbf{B}=0 \,.
\end{eqnarray}
Here ${\rm {\mathbb P}}\equiv p_{\perp}{\rm {\mathbb I}}+ \sigma_{d}{\bB}{\bB}/\mu_{0}$ is the CGL pressure tensor \cite{CGL}, where the function $\sigma_{d}\equiv \mu_{0}(p_{\parallel}-p_{\perp})/B^{2}$ measures pressure anisotropy. For incompressible flow,  under the assumption that $\sigma_d$ is uniform on magnetic surfaces  and the condition of helical symmetry, we have derived a  GGS equation  in helical coordinates $(r, \, u=m\phi -kz, \, z)$, where $(r, \, \phi \, z)$ are cylindrical coordinates \cite{evku18}:
\vspace{-2mm}
\begin{equation}
\label{GGS}
{\cal L}U
+\frac{2kmqX}{(1-\sigma_{d}-M_{p}^{2})^{1/2}}
+\frac{1}{2}
\left(\frac{X^{2}}{1-\sigma_{d}-M_{p}^{2}}\right)^{'}+\frac{\mu_{0}}{q}\bar{p}_{s}^{'}+
\frac{\mu_{0}}{2q^{2}}
\left[(1-\sigma_{d})\varrho(\Phi ^{'})^{2}\right]^{'}=0 \, .
\end{equation}
Here $U(r,u)=\int_{0}^{\psi}(1-\sigma_{d}(f)-M_{p}^{2}(f))^{1/2}df$, where $(2\pi m/k)\psi$ is the poloidal magnetic flux; $\varrho(U)$ is the mass density, $X(U)$ is related to the helical magnetic field, $M_p$ is the poloidal Mach function related to the  parallel component   of the flow, $\Phi(U)$ is the electrostatic potential in connection with the non-parallel component of the flow; $\bar{p}_s(U)$ is the static part of an effective pressure defined as $\bar{p}\equiv (p_{\perp}+p_{\parallel})/2$; $q\equiv (k^2r^2+m^2)^{-1}$ is a scale factor connected with the helical symmetry; ${\cal L}\equiv (1/q)\mathbf{\nabla}\cdot(q\mathbf{\nabla})$; and the prime denotes  derivative with respect to  $U$. The equilibrium is governed by (\ref{GGS}) and a Bernoulli equation for $\bar{p}$ \cite{evku18}. 
As an example,  in Fig. 1 we present 2-dimensional equilibria of Wendelstein 7x,  depicting features  in the limit of zero toroidicity and constant torsion,  constructed by an analytic solution of Eq. (\ref{GGS}).
\vspace{-1mm}
\begin{figure}[H]
\centering
\includegraphics[width=0.3\textwidth]{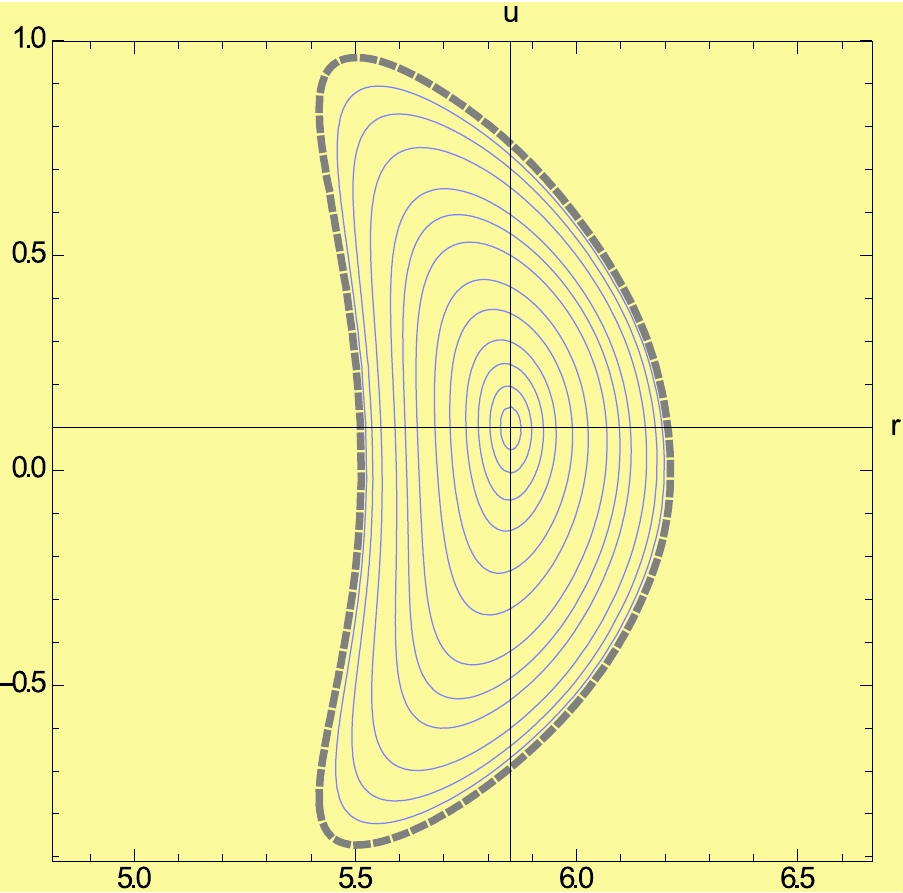}
\includegraphics[width=0.3\textwidth]{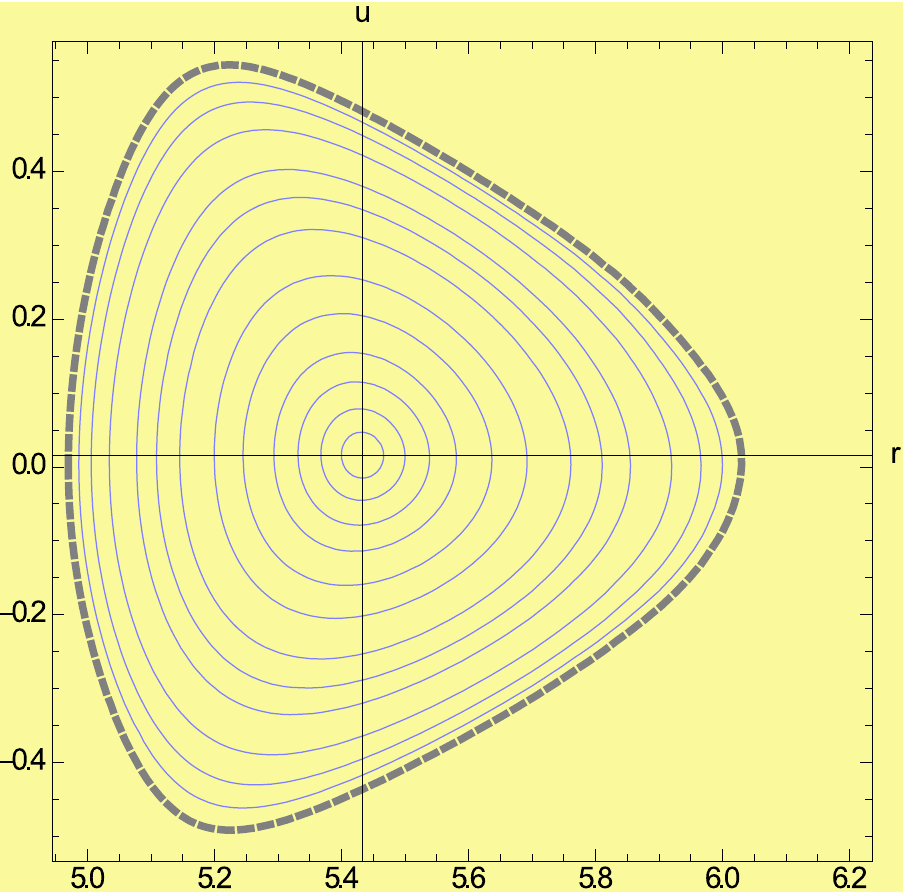}
\vspace{-1.5mm}
\caption{\emph{\footnotesize Poloidal cuts of the  magnetic surface which  remain invariant along the helical direction resembling  respective  ones of WD7-x stellarator for toroidal angles $0^o$ and $36^{o}$ \cite{Faustin}-\cite{Drevlak}.}}
\end{figure}
\vspace{-3.5mm}
\begin{figure}[H]
\centering
\includegraphics[width=0.32\textwidth]{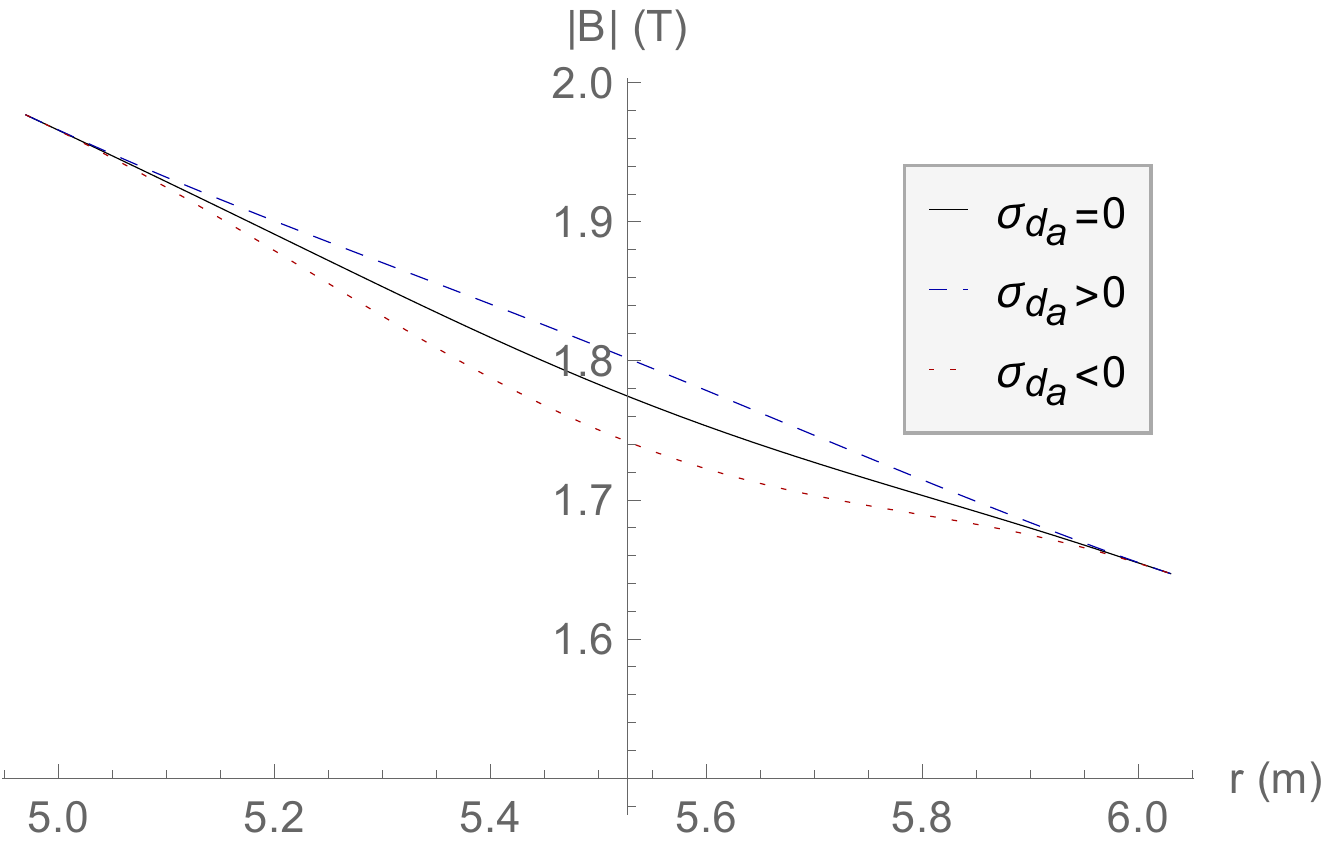}
\includegraphics[width=0.32\textwidth]{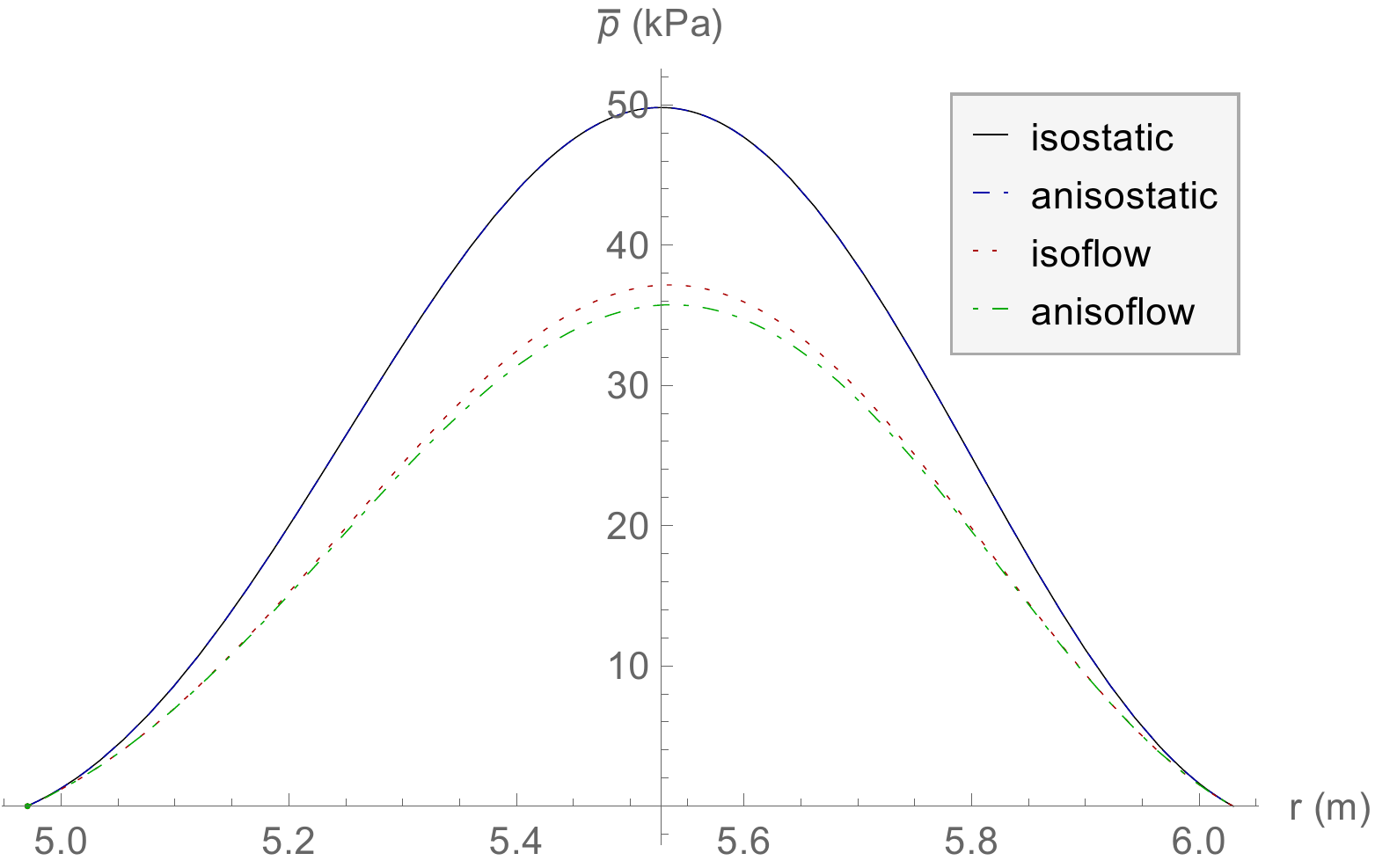}
\includegraphics[width=0.32\textwidth]{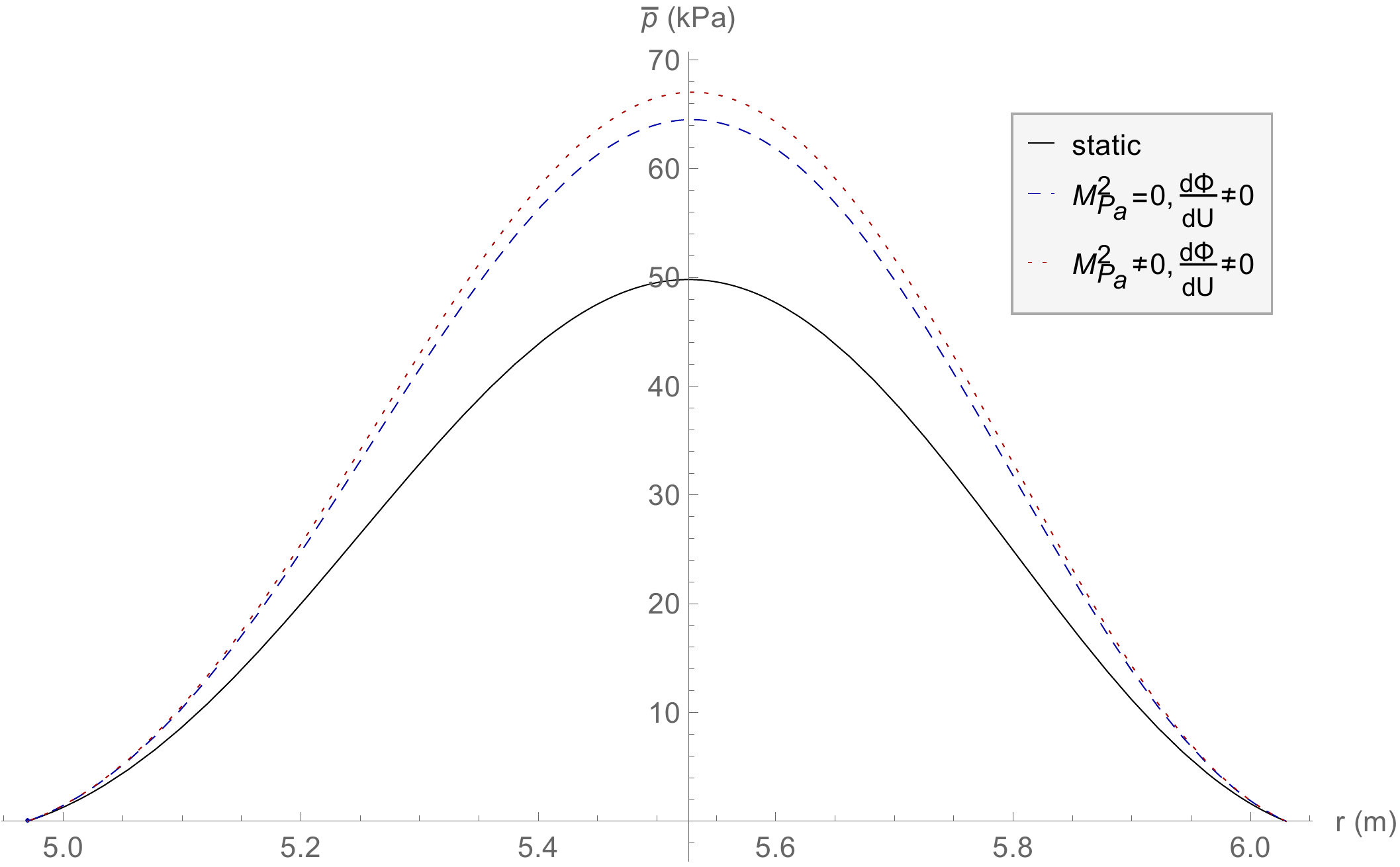}
\caption{\emph{\footnotesize The left figure shows the paramagnetic/diamagnetic impact of anisotropy. The central  figure  illustrates  the synergetic paramagnetic action of parallel flow and anisotropy in the absence of electric field. The right figure shows the diamagnetic impact of the electric field, which is enhanced by the parallel flow.}} \vspace{-2mm}
\end{figure}
Also, we  found that pressure anisotropy can act either  paramagnetically (for $\sigma _d>0$) or diamagnetically (for $\sigma _d<0$); note that $\sigma_d$ can be  either positive or negative depending on the direction of the auxiliary heating. For $\sigma_d>0$ magnetic-field-aligned flow has  an additive paramagnetic impact to that of anisotropy. The non-parallel flow has a diamagnetic impact, which becomes stronger as  $M_p^2$ takes larger values
 (Fig. 2).

{\bf Numerical axisymmetric equilibria with pressure anisotropy and parallel flow}:
For parallel flow Eq. (\ref{GGS}) becomes identical in form with the usual isotropic static Grad-Shafranov equation.
The impact of  plasma flow can be examined by means of the (total) Alfv\'enic Mach function, $M$,  which for parallel flow becomes identical with  the poloidal  Alfv\'enic Mach function, $M_p$ .
%
 Here  we have extended the axisymmetric equilibrium code HELENA \cite{Hu} to 
equilibria with pressure anisotropy and  parallel incompressible flow on the basis of the axisymmetric form of Eq. (\ref{GGS}) (for parallel flow). 
In this case the physical quantities are calculated by means of the 
aforementioned transformation, $U(r,z)=\int_{0}^{\psi}(1-\sigma_{d}(f)-M_{p}^{2}(f))^{1/2}df$,  in a manner similar to that employed for the extension of HELENA for 
parallel flow and isotropic pressure \cite{Pou}.  
The free functions $\sigma_d(U)$ and $M^2(U)$ which can be peaked on- or 
off-axis have been chosen  respectively as: 
\vspace{-3.5mm}
\begin{equation}
 F = F_0 \left(U^m - U_b^m\right)^n,\   F= C 
\left[\left(\frac{U}{ U_b}\right)^m\left( 
1-\left(\frac{U}{ U_b}\right)\right)\right]^n,
\end{equation}
where $F$ stands for  $M^2$ or $\sigma_d$; $m$, $n$ are shaping parameters;  and $C  =  F_0[(m+n)/m]^m [n/(m+n)]^n $.
 \begin{wrapfigure}{r}{113mm}
 \centering
\vspace{-0.65cm}
\includegraphics[scale=0.7, trim={8.9cm 9cm 9.5cm 
4cm},clip]{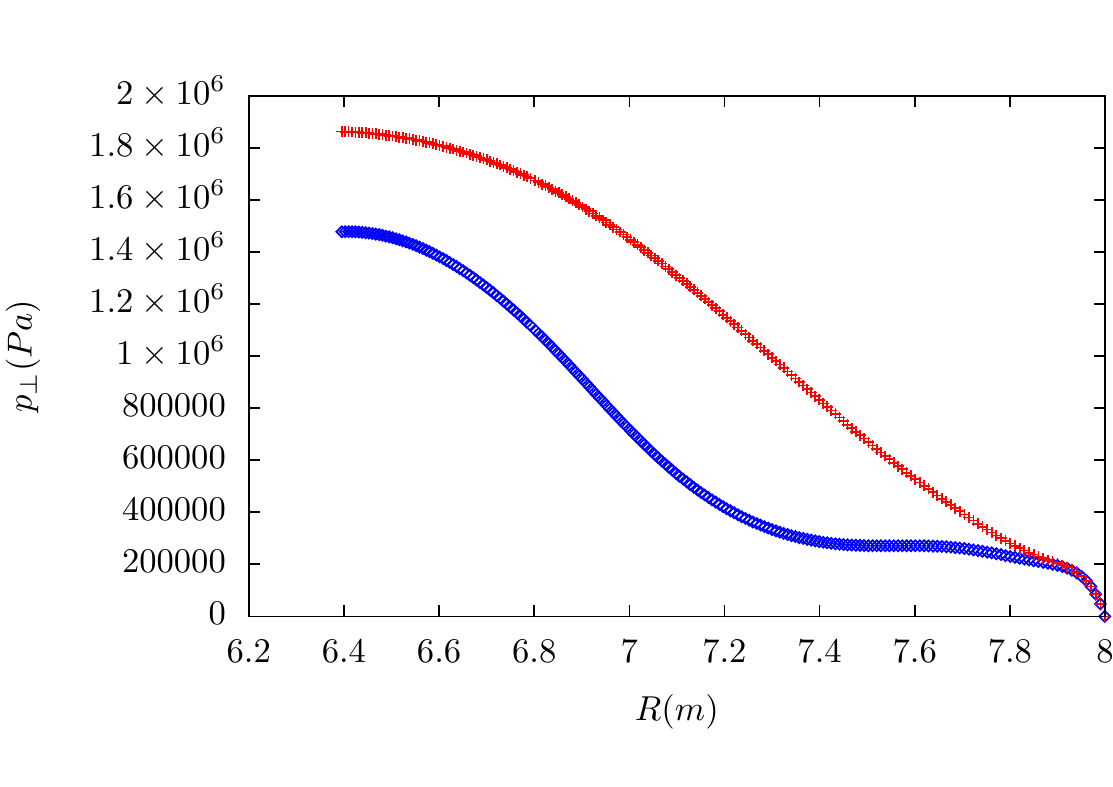}
\vspace{-1 cm}
\caption{{\emph{\footnotesize  Perpendicular   pressure profiles  for
 $M_0=0.04$, $m_M=5$, $n_M=2$, $\sigma_0=0.02$, $m_\sigma=2$, 
$n_\sigma=3$ (red curve);  and $M_0=0.04$, $m_M=5$, $n_M=2$, $\sigma_0=-0.02$, $m_\sigma=2$, 
$n_\sigma=3$ (blue curve).}}}
\label{fig:1}
\vspace{-0.58cm}
\end{wrapfigure}
The results indicate that the pressure anisotropy affects some quantities such 
as the current density or the magnetic field while for others, such as the 
effective pressure,  the impact of $\sigma_d$ is activated only in the presence of flow. In the latter case  the
impact of $\sigma_d$ is weak compared to that  caused by the flow. The 
presence of both pressure anisotropy and flow provides 
freedom in  the profile shaping  and results  in larger values of some equilibrium
quantities. For example, in the case of peaked on-axis $\sigma_d$ and peaked 
off-axis $M^2$ the shape of the perpendicular  pressure can be changed effectively  as 
can be seen in Fig. \ref{fig:1}.

{\bf Two-fluid effects}:
A  generalization of the aforementioned equilibrium  studies for isotropic pressure can be obtained  if in addition to  macroscopic flows one considers two-fluid effects.
\begin{wrapfigure}{r}{62mm}
\centering
\vspace{-6mm}
\includegraphics[scale=0.43]{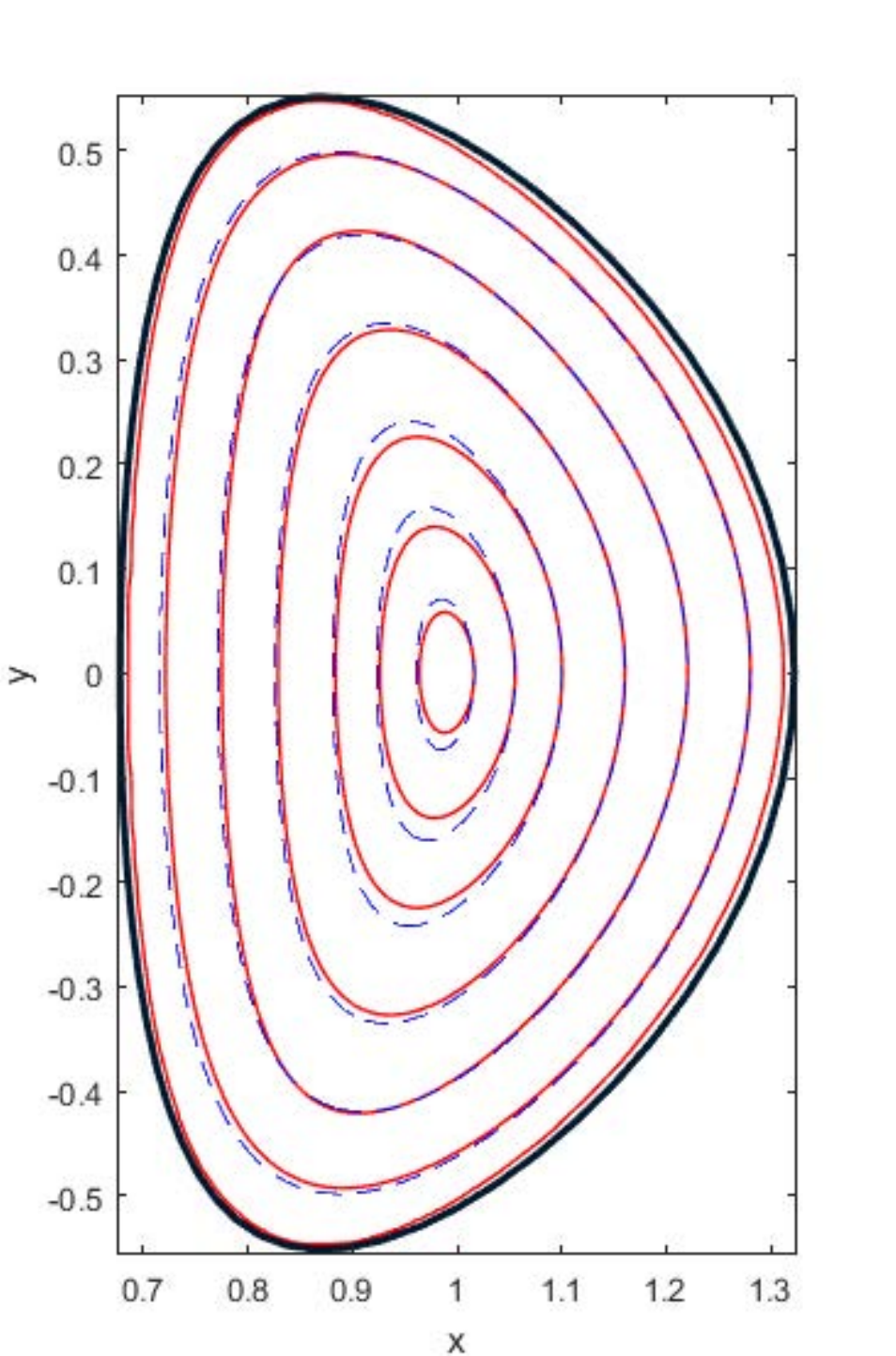} 
\vspace{-0.6cm}
\caption{\emph{\footnotesize  Ion flow surfaces
 (solid-red)  and magnetic surfaces
(dashed-blue) with 
$d_i = 0.03$ for a ``straight'' Tokamak HMHD equilibrium. The solid black line represents the boundary. }}
 \label{fig_hall}
\vspace{-0.6cm} 
\end{wrapfigure}
 In our study we employ  simplified versions of the complete two-fluid model  obtained by the imposition of the quasineutrality condition. 
Such models are the compressible, barotropic ($P=P(\rho)$) XMHD and the Hall MHD (HMHD). The former is obtained by expanding  in the smallness of the ratio $m_e/m_i$ and keeping up to first order terms and the latter by assuming massless electrons. The equilibrium studies were conducted within a Hamiltonian framework by constructing the corresponding helically symmetric Hamiltonian formulation, through the computation of the symmetric Hamiltonian functional and noncanonical Poisson bracket \cite{kaltsas2017,kaltsas2018}. Equilibrium equations were obtained through the Energy-Casimir variational principle. This principle allows for equilibrium and stability studies through the variation of the Hamiltonian functional with constraints being the various Casimir invariants of the model, which are functionals Poisson-commuting with any arbitrary functional $F$ defined in  the functional phase space. 
We applied this procedure for XMHD which is described by the following equations \vspace{-0.4cm}
\begin{eqnarray}
&&\partial_{t}\rho=-\nabla\cdot\left(\rho \bv\right), \ \ 
\partial_t \bv=\bv\times(\nabla\times\bv)-\nabla v^2/2-\rho^{-1}\nb p+\rho^{-1}\bJ\times\bB^*
  -d_e^2\nabla\left(\frac{|\bJ|^2}{2\rho^2}\right)\,, \label{mom_eq} \nonumber \\ [-2mm]
&&\partial_t\boldsymbol{B}^*=\nabla\times\left(\bv\times\bB^*\right)-d_i\nabla\times\left(\rho^{-1}\bJ\times \bB^*\right) +d_e^2\nabla\times\left[\rho^{-1}\bJ\times\left(\nabla\times\bv\right)\right] \,,
\label{starB}
\end{eqnarray}
where
$\bJ=\nabla\times\bB$, $\bB^*=\bB+d_e^2\nabla\times(\nb\times\bB/\rho)$ and the parameters $d_i$ and $d_e$ are  normalized ion and electron skin depths respectively. The HMHD system is obtained by $d_e=0$. We found that the symmetric versions of this model possess four families of Casimirs denoted by $\Cc_i$, $i=1,...,4$. Employing the Energy-Casimir equilibrium variational principle, $\delta(\Hc-\sum_{i=1}^4\Cc_i)=0$,  for the helically symmetric version,  we found a system  of equilibrium equations which can be cast in  the form of a Grad-Shafranov-Bernoulli (GSB) set,   consisting of three  coupled partial differential equations (PDEs) with respect to  the poloidal magnetic flux $\psi$, and two additional poloidal stream functions $\varphi=\psi^*+\gamma_+ q^{-1/2} v_h$ and  $\xi=\psi^*+\gamma_- q^{-1/2} v_h$; here  $\psi^*$ is the poloidal flux function of the generalized magnetic field $\bB^*$, $v_h$ is the component of the helical velocity, 
and $\gamma_\pm= [d_i\pm (d_i^2+4d_e^2)^{1/2}]/2$. Those PDEs are additionally coupled to a Bernoulli equation \cite{kaltsas2017,kaltsas2018}. Setting $d_e=0$ one obtains the Hall MHD GSB system. For reasons of conciseness we present here only the HMHD GSB;  the reader is referred to \cite{kaltsas2018} for the  complete helically symmetric XMHD system and to \cite{kaltsas2017} for its translationally symmetric counterpart. The HMHD GSB system of equations is
\vspace{-0.2cm}
\begin{eqnarray}
d_i^2\Fc'\nb\cdot\left(\frac{q}{\rho}\nb\Fc\right)=q(\Fc+\Gc)\Fc'+\rho \Mc'- q\left[\frac{\rho}{d_i^2}+2k m q\Fc'\right](\varphi-\psi)\,,\label{hall_jfko_1} \\ [-2.5mm]
\tilde{\Lc}\psi=q(\Fc+\Gc)\Gc'+\rho \Nc'+2k m q^2(\Fc+\Gc)+q \rho\frac{(\varphi-\psi)}{d_i^2}\,,\label{hall_jfko_2}\\ [-2mm]
h(\rho)=\left[\Mc+\Nc-q\frac{(\varphi-\psi)^2}{2d_i^2}\right]-d_i^2 q\frac{(\Fc')^2}{2\rho^2}|\nb\varphi|^2\,, \label{bernoulli_hall} 
\end{eqnarray}
where $\Fc$, $\Mc$  and  $\Gc$, $\Nc$ are arbitrary functions of $\varphi=\psi+d_iq^{-1/2}v_h$ and $\psi$ respectively. The operator $\tilde{\Lc}$ is $\tilde{\Lc}\equiv -q\Lc=-\nb\cdot(q\nb)$. The system \eqref{hall_jfko_1}-\eqref{bernoulli_hall} is elliptic for subsonic poloidal flows and becomes hyperbolic for $v_p^2>c_s^2$. We computed numerically an  HMHD equilibrium state by solving the above system, in the subsonic regime, assuming translational symmetry ($k=0$). The resulting equilibrium configuration,  obtained by second-order polynomial ansatzes for the free functions, is depicted in Fig. \ref{fig_hall}. We observe that the flow surfaces depart from the magnetic ones,  as expected  in  the framework of  HMHD model.  Determining the separation distance of the two sets of characteristic surfaces may be of interest for transport studies.

{\bf Acknowledgments}: {\footnotesize This work has been carried out within the 
framework of the EUROfusion
Consortium and has received funding from (i) the National Programme for the
Controlled Thermonuclear Fusion, Hellenic Republic and (ii) Euratom research
and training program 2014-2018 under grant agreement no. 633053. The views
and opinions expressed herein do not necessarily reflect those of the European
Commission. A.E. and D.A.K. were  supported by  PhD grants from the Hellenic Foundation
for Research and Innovation (HFRI) and the General Secretariat for Research and
Technology (GSRT).
P.J.M. was supported by the US Department of Energy contract DE-FG05-
80ET-53088 and a Forschungspreis from the Alexander von Humboldt Foundation.}

\end{document}